\documentclass{PoS}
\usepackage{axodraw}
\usepackage{cite}
\newcommand{\bqa}{\begin{eqnarray}}
\newcommand{\eqa}{\end{eqnarray}}
\newcommand{\nl}{\nonumber \\}
\newcommand{\dbar}{\bar D}

\newcommand{\qbar}{\bar q}

\title{A fresh look at (non)renormalizable QFTs}

\ShortTitle{A fresh look at (non)renormalizable QFTs}
\author{\speaker{Roberto Pittau}\thanks{Work performed in the framework of the ERC grant 291377 (LHCtheory), the MICINN project FPA2011-22398 (LHC@NLO) and the Junta de Andaluc\'ia project P10-FQM-6552.}\\
        Departamento de F\'isica Te\'orica y del Cosmos and CAFPE,
  Campus Fuentenueva s. n., Universidad de Granada, E-18071 Granada, Spain \\
        E-mail: \email{pittau@ugr.es}}

\abstract{Abandoning dimensional regularization allows important simplifications in loop calculations and gives a handle to interpret non-renormalizable 
Quantum Field Theories. I review the current status of FDR, a fully four-dimensional approach to the ultraviolet problem.}

\FullConference{11th International Symposium on Radiative Corrections (Applications of Quantum Field Theory to Phenomenology) (RADCOR 2013),\\
		22-27 September 2013\\
		Lumley Castle Hotel, Durham, UK }

\begin{document}
\section{Introduction}
Although renormalizability provides a powerful guiding principle when searching for fully consistent Quantum Field Theories (QFT) at the fundamental level, it may very well be that not all interactions admit a microscopic description in terms of renormalizable or free-of-infinities theories. On the other hand, the physics content of the non-renormalizable QFTs is much reacher, since operators of higher dimensionality are allowed, that can be useful in the effective description of physical phenomena at energies at which the {\em true} theory is unknown (or non-perturbative).

Non-renormalizable theories are normally dealt with within the same framework of the renormalizable ones at the price of giving up a bit of predictivity at each additional perturbative order: operators needed to re-absorb infinities generated by the new virtual loops are introduced at the fundamental level, that require to be fixed by experimental observables. Then the theory is order-by-order predictive, but an increasingly large number of data points is needed, and predictivity is totally lost in the infinite loop limit. 

In this contribution, I illustrate an alternative procedure: since an infinite energy is required to resolve a vanishing space-time distance between two points, it is reasonable to assume that all ultraviolet (UV) infinities are non-physical/unobservable degrees of freedom  - generated by the loop expansion -  that have to be separated from the physical spectrum. The UV problem can then be recast as the problem of separating infinities and physics in an unambiguous way, respecting - at the same time - the symmetries of the Lagrangian.
Such a separation can be naturally obtained within the FDR approach of~\cite{Pittau:2012zd}, which allows to
\begin{itemize}
\item reproduce the physics of the renormalizable theories (bottom-up approach);
\item give a sensible meaning to the non-renormalizable QFTs (top-down approach).
\end{itemize}
Known one-loop~\cite{Donati:2013iya,Pittau:2013qla} and two-loop~\cite{Donati:2013voa} results have been recently re-derived in FDR, showing its consistency and correctness in the bottom-up direction.
In the next section, I review the basic features of FDR, while section~\ref{topdown} describes how FDR works in the non-renormalizable case.
\section{Bottom-up}
\subsection{FDR integration}
In FDR, the UV infinities are subtracted at the {\em integrand} level by judiciously splitting the original integrand $J(q_1,\ldots, q_\ell)$ of an $\ell$-loop function~\footnote{$q_1,\ldots, q_\ell$ are integration momenta and $J(q_1,\ldots, q_\ell)$ can be a tensor.} in two parts, 
$J_{\rm INF}(q_1,\ldots, q_\ell)$ and $J_{{\rm F},\ell}(q_1,\ldots, q_\ell)$. The former piece collects integrands which would give divergences upon integration, while the latter generates the finite contribution. To avoid the occurrence of infrared divergences the $+i 0$ propagator prescription has to be made explicit by identifying it with a vanishing mass $-\mu^2$ and taking the limit $\mu \to 0$ outside integration. The rationale for this separation is that the loop {\em integrands} in $J_{\rm INF}(q_1,\ldots, q_\ell)$ are allowed to depend on $\mu$, {\em but not on physical scales}, so that the physics is entirely contained in $J_{{\rm F},\ell}(q_1,\ldots, q_\ell)$. As an explicit two-loop example consider
\bqa
J^{\alpha \beta}(q_1,q_2) = \frac{q^\alpha_1 q^\beta_1}{\dbar^3_1 \dbar_2 \dbar_{12}}\,,
\eqa
with
\bqa
\label{eq:2lden}
\bar D_1   = \bar{q}_1^2-m_1^2\,,~~~~
\bar D_2   = \bar{q}_2^2-m_2^2\,,~~~~
\bar D_{12} = \bar{q}_{12}^2-m_{12}^2\,,~~~~
q_{12}= q_1+q_2\,,~~\bar q^2_j = q^2_j-\mu^2\,. 
\eqa
The needed splitting can be obtained through a repeated use of the identities
\bqa
\label{eq:ids}
\frac{1}{\dbar_j}  = \frac{1}{\qbar^2_j} +\frac{m^2_j}{\qbar^2_j \dbar_j}\,,~~~~
\frac{1}{\qbar^2_{12}}= \frac{1}{\qbar^2_2}-\frac{q^2_1+2 (q_1 \cdot q_2)}{\qbar^2_2 \qbar^2_{12} }\,,~~~~
\frac{1}{\qbar^2_{2}}= \frac{1}{\qbar^2_1}-\frac{q^2_{12}-2 (q_1 \cdot q_{12})}{\qbar^2_1 \qbar^2_2 }\,,
\eqa
and reads
\bqa
\label{eq:appa1}
J^{\alpha \beta}(q_1,q_2) &=&
q^\alpha_1 q^\beta_1 
\left\{
\left[\frac{1}{\qbar^6_1 \qbar^2_2 \qbar^2_{12}}  \right]
  +\left(
        \frac{1}{\bar D^3_1}
       -\frac{1}{\bar{q}_1^6}
    \right)
\left(
 \left[\frac{1}{\qbar_2^4}\right]
 -\frac{q_1^2+2(q_1 \cdot q_2)}{\bar{q}_2^4 \bar{q}_{12}^2} 
\right)
\right. \nl
 &&+\left.\frac{1}{\bar D_{1}^3 \bar{q}_2^2\bar D_{12} }
 \left(
 \frac{m_2^2}{\bar D_{2}}+\frac{m_{12}^2}{\bar{q}_{12}^2}
 \right) 
\right \}\,,
\eqa
where divergent integrands are written between square brackets.
Then
\bqa
\label{eq:ex2l}
J^{\alpha \beta}_{\rm INF}  (q_1,q_2) &=& 
q^\alpha_1 q^\beta_1 
\left\{
\left[\frac{1}{\qbar^6_1 \qbar^2_2 \qbar^2_{12}}  \right]
  +\left(
        \frac{1}{\bar D^3_1}
       -\frac{1}{\bar{q}_1^6}
    \right)
 \left[\frac{1}{\qbar_2^4}\right]
\right \}\,~~~{\rm and} \nl
J^{\alpha \beta}_{{\rm F},2}(q_1,q_2) &=& 
q^\alpha_1 q^\beta_1 
\left\{
 \frac{1}{\bar D_{1}^3 \bar{q}_2^2\bar D_{12} }
  \left(
  \frac{m_2^2}{\bar D_{2}}+\frac{m_{12}^2}{\bar{q}_{12}^2}
  \right) 
  -\left(
        \frac{1}{\bar D^3_1}
       -\frac{1}{\bar{q}_1^6}
    \right)
 \frac{q_1^2+2(q_1 \cdot q_2)}{\bar{q}_2^4 \bar{q}_{12}^2} 
\right\}\,.
\eqa 
The FDR integral over the original integrand $J(q_1,\ldots, q_\ell)$ is {\em defined} as~\footnote{FDR integration is denoted by the symbol $[d^4q_i]$.}
\bqa
\label{eq:fdrdef}
\int [d^4q_1] \ldots [d^4q_\ell]\,  J(q_1,\ldots, q_\ell) \equiv \lim_{\mu \to 0}
\int d^4q_1 \ldots d^4q_\ell\, J_{{\rm F},\ell}(q_1,\ldots q_\ell)\,,
\eqa
and the expansion needed to extract $J_{{\rm F},\ell}(q_1,\ldots q_\ell)$ is called
the {\em FDR defining expansion} of $J(q_1,\ldots, q_\ell)$.
For example, from eq.~(\ref{eq:ex2l}),
\bqa
\int [d^4q_1] [d^4q_2] \frac{q^\alpha_1 q^\beta_1}{\dbar^3_1 \dbar_2 \dbar_{12}}
= \lim_{\mu \to 0} \int d^4q_1 d^4q_2\,J^{\alpha \beta}_{{\rm F},2}(q_1,q_2)\,.
\eqa
The advantage of using FDR integration in gauge QFTs is that it encodes the UV subtraction {\em directly} into its definition, maintaining, at the same time, the two properties needed to prove Ward Identities, i.e.
\begin{itemize}
\item[i) ] invariance under shift of any integration variable;
\item[ii)] simplifications among numerators and denominators.
\end{itemize}
The first property follows by rewriting FDR integrals as a finite
differences of UV divergent integrals:
\bqa
\label{eq:diff}
\int [d^4q_1] \ldots [d^4q_\ell]\,  J(q_1,\ldots, q_\ell) = 
\lim_{\mu \to 0} 
 \int d^nq_1 \ldots d^nq_\ell \,  
\Big(
J(q_1,\ldots, q_\ell) - J_{\rm INF}(q_1,\ldots, q_\ell)
\Big)\,,
\eqa 
where the r.h.s. is regulated in dimensional regularization~\footnote{This is just one option, since the dependence on {\em any} UV regulator drops in the difference.} (DR). The second property is guaranteed by construction, provided any $q^2_i$ generated by Feynman rules is considered as $\qbar^2_i= q^2_i-\mu^2_i$~\footnote{Only one kind of $\mu^2$ exists. The index $_i$ in $\mu^2_i$ only denotes that the denominator expansion in front of $\mu^2$ should be {\em the same one used for $q^2_i$}.}. For example, from the defining expansions of the three integrands:
\bqa
\label{2loopFDR}
\int [d^4q_1] [d^4q_2] 
  \frac{\qbar^2_1}{\bar D_1^3\bar D_2\bar D_{12}} &=&
\int [d^4q_1] [d^4q_2] 
  \frac{1}{\bar D_1^2\bar D_2\bar D_{12}} +
m_1^2 \int [d^4q_1] [d^4q_2] 
  \frac{1}{\bar D_1^3\bar D_2\bar D_{12}}\,.
\eqa
For consistency, tensor decomposition works as in the following example
\bqa
\!\!\!\!\!\!\int [d^4q_1] [d^4q_2] 
  \frac{q^\alpha_1q^\beta_1}{\bar D_1^3\bar D_2\bar D_{12}} 
=
\frac{g^{\alpha \beta}}{4}\!\!
\int [d^4q_1] [d^4q_2] 
 \frac{q_1^2}{\bar D_1^3\bar D_2\bar D_{12}} 
=
\frac{g^{\alpha \beta}}{4}\!\!
\int [d^4q_1] [d^4q_2] 
 \frac{\qbar_1^2+\mu^2_1}{\bar D_1^3\bar D_2\bar D_{12}}\,, 
\eqa
where, from eq.~(\ref{eq:ex2l}),
\bqa
\int [d^4q_1] [d^4q_2] 
 \frac{\mu^2_1}{\bar D_1^3\bar D_2\bar D_{12}}= 
\lim_{\mu \to 0}\, \mu^2 \int d^4q_1 d^4q_2 
\frac{g_{\alpha\beta}\,J^{\alpha \beta}_{{\rm F},2}(q_1,q_2)}{q^2_1}\,.
\eqa
It is interesting to investigate how FDR integrals depend on $\mu$~\footnote{In the absence of IR divergences.}.
The first term in the r.h.s. of eq.~(\ref{eq:diff}) does not depend on $\mu$, because $\lim_{\mu \to 0}$ can be moved inside integration. On the other hand, any polynomially divergent integral in $J_{\rm INF}(q_1,\ldots, q_\ell)$ cannot contribute either, being proportional to positive powers of $\mu$.  Therefore, the $\mu$ dependence of the l.h.s. of eq.~(\ref{eq:diff}) is entirely due to powers of $\ln (\mu/\mu_R)$~\footnote{$\mu_R$ is the arbitrary renormalization scale of DR, that can also be thought as the arbitrary scale at which one decides to subtract the logarithmically divergences.} generated by the logarithmically divergent subtracted integrals.
Therefore:
\begin{itemize}
\item[i) ] FDR integrals depend on $\mu$ {\em logarithmically};
\item[ii) ] if all powers of $\ln (\mu/\mu_R)$ are moved to the l.h.s. of eq.~(\ref{eq:diff})~\footnote{From now on, the FDR integration is redefined by assuming this.}, the $\lim_{\mu \to 0}$ can be taken by formally trading $\ln (\mu)$ for
$\ln (\mu_r)$. 
\end{itemize}
Then, FDR integrals {\em do not depend on any cut off} but only on the renormalization scale $\mu_R$.

\subsection{Infrared and collinear infinities}
The FDR treatment of the UV infinities is compatible with the presence of Infrared (IR) or collinear (CL) divergences in massless theories.
The basic observation is that the $+i0= -\mu^2$ prescription naturally regulates any IR/CL behavior in the loop integrals:
IR/CL divergent loop integrals are unambiguously defined by taking the limit $\mu \to 0$ outside integration - as in the UV case - after subtracting divergent integrands, when necessary. This should be matched with a consistent treatment of IR/CL infinities generated in the real part of the calculation, that corresponds to a massless calculation of the real matrix element squared $|M|^2$ integrated over a phase-space where all would-be-massless particles are given a mass 
$\mu$, with $\mu \to 0$. This can be understood because 
\begin{itemize}
\item a massless calculation of $|M|^2$ preserves gauge invariance;
\item unitarity relates the $+i0= -\mu^2$ deformation in a massless $1/\qbar^2$ loop propagator to a phase-space integration with $q^2= \mu^2$.    
\end{itemize}
Such a procedure has been proven to work at one loop in~\cite{Pittau:2013qla}.  
\subsection{Renormalization in FDR}
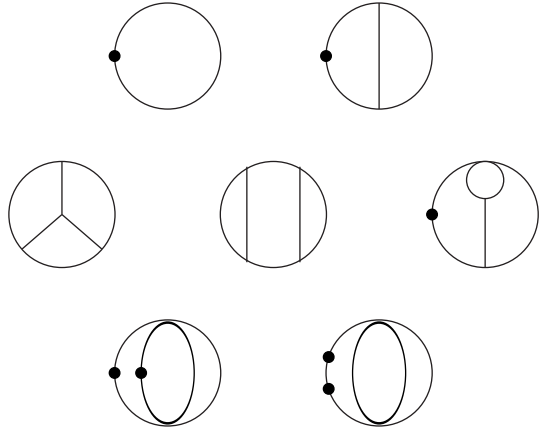
\begin{figure}[t]
\begin{center}
\begin{picture}(300,160)(0,0)
\SetOffset(140,130)
\BCirc(-30,12){20}
\CCirc(-50,12){2}{}{}

\SetOffset(220,130)
\BCirc(-30,12){20}
\CCirc(-50,12){2}{}{}
\Line(-30,32)(-30,-8)

\SetOffset(100,70)
\BCirc(-30,12){20}
\Line(-30,32)(-30,12)
\Line(-30,12)(-15,-1)
\Line(-30,12)(-45,-1)

\SetOffset(180,70)
\BCirc(-30,12){20}
\Line(-40,30)(-40,-6)
\Line(-20,30)(-20,-6)

\SetOffset(260,70)
\BCirc(-30,12){20}
\CCirc(-50,12){2}{}{}
\Line(-30,18)(-30,-8)
\BCirc(-30,25){7}

\SetOffset(140,10)
\BCirc(-30,12){20}
\CCirc(-50,12){2}{}{}
\GOval(-30,12)(10,19)(90){1}
\CCirc(-40,12){2}{}{}

\SetOffset(220,10)
\BCirc(-30,12){20}
\GOval(-30,12)(10,19)(90){1}
\CCirc(-49,18){2}{}{}
\CCirc(-49,6){2}{}{}

\end{picture}
\end{center}
\caption{Logarithmic scalar topologies (up to three loops) subtracted by FDR. Tensors are also reducible to them. Dotted and double-dotted propagators are raised by one and two additional powers, respectively.}
\label{fig:fig1}
\end{figure}

In FDR any calculation is UV finite by construction. Nevertheless it is illuminating to consider the subtraction embedded in the definition of FDR integral as the operation of redefining the vacuum: order by order in the perturbation theory observable physics is defined with respect to a new vacuum, obtained by subtracting the unphysical divergent configurations (which can be interpreted as {\em vacuum bubbles}) contained in $J_{\rm INF}(q_1,\ldots, q_\ell)$. All polynomially divergent vacuum bubbles can be subtracted {\em at no price}, while the logarithmic ones, classified in figure~\ref{fig:fig1} up to three loops,  leave logarithms of $\mu_R$.    
This interpretation, dubbed {\em Topological Renormalization}~\cite{Pittau:2013ica}, differs from the usual renormalization procedure - used, for example, in DR - in that the Lagrangian ${\cal L}$ is left {\em untouched} and no counterterms need to be added. What changes order by order is not ${\cal L}$, but the vacuum, so that the parameters $p_i$ ($i= 1:m$) upon which ${\cal L}(p_1,\ldots,p_m)$ depends - i.e. couplings and masses - remain finite. However, they still have to be linked to experimental observables by means of a finite renormalization. In particular, $m$ measurements 
\bqa
{\cal O}_{i}^{\rm TH}(p_1,\ldots,p_m) = {\cal O}_{i}^{\rm EXP}
\nonumber
\eqa 
are needed to determine {$p_i$} in terms of observables {${\cal O}^{\rm EXP}_{i}$} and corrections computed at the loop level { $\ell$} one is working:
\bqa
p_i= p^{\ell-loop}_i({\cal O}^{\rm EXP}_{1},\ldots,  {\cal O}^{\rm EXP}_{m}) \equiv \bar p_i\nonumber
\eqa
Then
\bqa
{\cal O}_{m+1}^{\rm TH}(\bar p_1,\ldots,\bar p_m)\, \nonumber
\eqa
is a finite prediction of the QFT and, if the theory is renormalizable, the dependence on $\mu_R$ drops:
\bqa
\label{eq:der}
\frac{\partial {\cal O}_{m+1}^{\rm TH}(\bar p_1,\ldots,\bar p_m)}{\partial \mu_R} = 0\,.
\eqa

The absence of counterterms simplifies FDR loop calculations. Consider, for example~\cite{'tHooft:1973pz}, the one-loop photon self-energy in QED
\begin{center}
  \begin{picture}(200,50)(0,0)
    \SetOffset(-70,20)
    \Text(-18,-7)[t]{$\alpha$}
    \Text(-9, 12)[b]{$p$}
    \SetScale{0.75}
    \LongArrow(-17,10)(-8,10)
    \SetScale{1}
    \Photon(-20,0)(0,0){2.5}{4}
    \ArrowArc(10,0)(10,180,0)
    \ArrowArc(10,0)(10,0,180)
    \Photon(20,0)(40,0){2.5}{4}
    \Text(38,-7)[t]{$\beta$}
    \Text(47,0)[l]{$=\,i\, T_{\alpha\beta}\, \Pi(p^2)\,,$}
    \Text(130,0)[l]{$T_{\alpha\beta} = g_{\alpha\beta}p^2-p_\alpha p_\beta \,,$}
    \Text(240,0)[l]{$\Pi(p^2)     =  
\frac{1}{\epsilon}\, \Pi_{-1} + \Pi_{0} + \epsilon\, \Pi_{1} \,.$}
  \end{picture}
\end{center}
In DR, the corresponding two-loop computation requires the addition of one-loop couterterms such that
\begin{center}
  \begin{picture}(200,50)(0,0)
    \SetOffset(0,20)
    \Photon(-20,0)(0,0){2.5}{4}
    \ArrowArc(10,0)(10,180,0)
    \ArrowArc(10,0)(10,0,180)
    \Photon(20,0)(40,0){2.5}{4}
    \Text(47,0)[l]{$+$}
    \Photon(60,0)(80,0){2.5}{4}
    \Photon(80,0)(100,0){2.5}{4}
    \Text(80,0)[]{\LARGE $\bullet$}
    \Text(107,0)[l]{$=\,i\, T_{\alpha\beta}\, \Pi_0 + {\cal O}(\epsilon)\,.$}
  \end{picture}
\end{center}
Therefore, at two loops,
\begin{center}
  \begin{picture}(200,50)(0,0)
    \SetOffset(-80,20)
    \Photon(-10,0)(0,0){2.5}{2}
    \ArrowArc(10,0)(10,180,0)
    \ArrowArc(10,0)(10,0,180)
    \Photon(20,0)(30,0){2.5}{2}
    \ArrowArc(40,0)(10,180,0)
    \ArrowArc(40,0)(10,0,180)
    \Photon(50,0)(60,0){2.5}{2}
    \Text(70,0)[]{$+$}
    \Photon(80,0)(90,0){2.5}{2}
    \ArrowArc(100,0)(10,180,0)
    \ArrowArc(100,0)(10,0,180)
    \Photon(110,0)(120,0){2.5}{2}
    \Text(120,0)[]{\LARGE $\bullet$}
    \Photon(120,0)(130,0){2.5}{2}
    \Text(140,0)[]{$+$}
    \Photon(150,0)(160,0){2.5}{2}
    \Text(160,0)[]{\LARGE $\bullet$}
    \Photon(160,0)(170,0){2.5}{2}
    \ArrowArc(180,0)(10,180,0)
    \ArrowArc(180,0)(10,0,180)
    \Photon(190,0)(200,0){2.5}{2}
    \Text(210,0)[]{$+$}
    \Photon(220,0)(230,0){2.5}{2}
    \Text(230,0)[]{\LARGE $\bullet$}
    \Photon(230,0)(245,0){2.5}{3}
    \Text(245,0)[]{\LARGE $\bullet$}
    \Photon(245,0)(255,0){2.5}{2}
    \Text(265,0)[l]{$= \,i\, T_{\alpha\beta}\, \Pi^2_0 + {\cal O}(\epsilon)\,.$}
  \end{picture}
\end{center}
In FDR, the product of two one-loop diagrams is simply the product of the two finite parts, so that one directly obtains
\begin{center}
  \begin{picture}(200,50)(0,0)
    \SetOffset(20,20)
    \Photon(-10,0)(0,0){2.5}{2}
    \ArrowArc(10,0)(10,180,0)
    \ArrowArc(10,0)(10,0,180)
    \Photon(20,0)(30,0){2.5}{2}
    \ArrowArc(40,0)(10,180,0)
    \ArrowArc(40,0)(10,0,180)
    \Photon(50,0)(60,0){2.5}{2}
    \Text(70,0)[l]{$= \,i\, T_{\alpha\beta}\, \Pi^2_{\rm FDR}(p^2)\,,$}
  \end{picture}
\end{center}
with $\Pi_{\rm FDR}(p^2)= \Pi_0$.
The previous example also shows that $\ell$-loop integrals are directly re-usable in ($\ell$+1)-loop calculations. For instance, the two-loop factorizable FDR integral
\bqa
	\int \frac{[d^4q_1]}{(\qbar_1^2-m^2_1)^{\alpha}}
	\times
	\int \frac{[d^4q_2]}{(\qbar_2^2-m^2_2)^{\beta}} 
\eqa
is simply the product of two one-loop FDR integrals. That {\em is not} the 
case in DR, where further expanding in $\epsilon$ is required.

\section{Top-down}
\label{topdown}
It is interesting to extend the FDR framework to a non-renormalizable QFT
described by a Lagrangian ${\cal L}_{NR}$. Things remain unchanged up to eq.~(\ref{eq:der}), which - in general - is no longer true. Therefore
\bqa
{\cal O}_{m+1}^{\rm TH}(\bar p_1,\ldots,\bar p_m,\ln \mu_R)\, \nonumber
\eqa
may depend on the arbitrary scale $\mu_R$.
However, if a particular combination of observables is constructed, in which $\mu_R$ disappears, it can still be unambiguously predicted by ${\cal L}_{NR}$.
In principle this can be achieved with {\em just one} additional observable - ${\cal O}_{m+2}^{\rm EXP}$ - by solving the equation
\bqa
\label{eq:mp2}
{\cal O}_{m+2}^{\rm TH}(\bar p_1,\ldots,\bar p_m, \ln \mu^\prime_R) = {\cal O}_{m+2}^{\rm EXP}\,,
\eqa
and setting $\mu_R= \mu^\prime_R $ in ${\cal O}_{m+1}^{\rm TH}$.
Notice that it is crucial the fact that, in FDR, the original cut-off $\mu \to 0$ is traded with an adjustable scale $\mu_R$.
In addition, one has to assume that the solution of eq.~(\ref{eq:mp2}) still allows a perturbative treatment, i.e.
\bqa
|g^2\ln \mu^\prime_R| < 1\,,
\eqa
where $g$ is the coupling constant of the theory.

The outlined strategy is rather new and it has not been verified in practice, yet. More investigation is needed, the first obvious case study being a non-renormalizable theory for which the renormalizable counterpart is known (such as four-fermion contact interactions vs the electroweak standard model).
Finally, it is worth mentioning that one is free to consider a theory described by ${\cal L}_{NR}$ as an effective one. The difference of FDR with respect to the traditional approach to non-renormalizable theories is a gain in predictivity: the standard way of absorbing infinities into the parameters of ${\cal L}_{NR}$ {\em forces} a change in its form - at higher perturbative orders - such that new fundamental interactions have to be fixed in terms of experimental measurements. On the contrary, no change in ${\cal L}_{NR}$ is required by FDR~\footnote{Of course, one might still need to add interactions to reproduce experimental data.}, and the extra measurement in eq. (\ref{eq:mp2}) is all one needs to fix the theory. The meaning of this measurement is disentangling the effects of the unknown UV completion of ${\cal L}_{NR}$ - parametrized with a logarithmic dependence on $\mu_R$ - from the physical spectrum.

\section{Conclusions}
FDR can be used as an easier approach to higher order calculations in QFTs.
It is simpler than DR because:
\begin{itemize} 
\item order-by-order renormalization is avoided;
\item a finite renormalization is only required to fix the parameters of the theory in terms of experimental observables;
\item $\ell$-loop integrals are directly re-usable in ($\ell$+1)-loop calculations, with no need of further expanding in $\epsilon$.
\end{itemize} 
In addition, infrared and collinear divergences can be dealt with within the same four-dimensional framework used to cope with the ultraviolet infinities. 

FDR also allows a novel interpretation of non-renormalizable theories in which
predictivity is restored. The basic idea is that infinities are unphysical and can be separated - in a gauge invariant way - from the physical spectrum. The remnant of this operation is a dependence on the renormalization scale in physical observables, which, however, can be fitted via one extra measurement. 
Explicit calculations in non-renormalizable QFTs are needed to consolidate this interpretation.

\bibliography{proc}{}
\bibliographystyle{JHEP}

\end{document}